\documentclass[11pt]{article}

\begin{document}

\title{Sewing string tree vertices with ghosts using canonical forms}

\author{Leonidas Sandoval Junior\thanks{E-mail address: dma2lsj@dcc.fej.udesc.br}\\ Department of Mathematics\\ Centro de Ci\^encias Tecnol\'ogicas\\ UDESC - Universidade do Estado de Santa Catarina - Brazil}

\maketitle 

\begin{abstract}
We effectively sew two vertices with ghosts in order to obtain a third, composite vertex in the most general case of cycling transformations. In order to do this, we separate the vertices into two parts: a bosonic oscillator part and a ghost oscillator part and write them as canonical forms.
\end{abstract}

\begin{flushright}\noindent PACS number: 11.25\end{flushright}

\vskip 0.5 cm 

\section{Introduction}

Sewing of vertices in order to build larger ones is a technique that has been used since the early days of String Theory. In \cite{ls1}, it has been shown how to use the Group Theoretic approach to String Theory \cite{pw1} in order to perform the sewing of two string vertices taking into account their ghost structure. Here we shall perform the sewing of two vertices with ghosts using another procedure, writing the two original vertices in terms of canonical forms (see the appendix) and then performing the sewing explicitly. This method has been used in \cite{cg1} and \cite{cg2} in order to build multiloop string amplitudes for the bosonic string. This is done in order to check the result obtained in \cite{ls1} and also to present an example of the use of canonical forms in the sewing of vertices.

The sewing procedure consists on taking two distinct vertices $U_1$ and $U_2^\dagger $, the first with $N_1$ strings and the second with $N_2$ strings, and choosing one string from each vertex to be sewn together. We shall be choosing leg $E$ from vertex $U_1$ and leg $F$ from vertex $U_2$. The sewn legs give rise to what we call propagator ${\cal P}$. This procedure is well explained in \cite{pw2} and \cite{ls1}.

In order to perform the sewing, we shall divide the so called physical vertex \cite{pw3} of an open bosonic string into a bosonic and a fermionic part. This is the vertex that has the correct number of ghosts. It can be written as
\begin{eqnarray}
U & = & \left( \prod_{\scriptstyle i=1\atop \scriptstyle i\neq E}^{N}{ }_i\langle 0|\right) \exp \left[ -\frac{1}{2}\sum_{\scriptstyle i,j=1\atop \scriptstyle i\neq j}^{N_1}\sum_{n,m=0}^\infty a^{\mu i}_nD_{nm}\left( \Gamma V^{-1}_iV_j\right) a^j_{m\mu }\right] \exp \left[ \sum_{\scriptstyle i,j=1\atop \scriptstyle i\neq j}^N\sum_{n=2}^\infty \sum_{m=-1}^\infty c_n^iE_{nm}\left( \Gamma V^{-1}_iV_j\right) b_m^j\right] \nonumber \\ 
\label{eq1}
 & & \times \prod_{r=-1}^1\sum_{j=1}^N\sum_{s=-1}^1E_{rs}(V_j)b_s^j\times \prod_{\scriptstyle i=1\atop \scriptstyle i\neq a,b,c}^N\sum_{j=1}^N\sum_{n=-1}^\infty e_n^{ij}b_n^j\ ,
\end{eqnarray}
where $a^i_{n\mu }$ are bosonic oscillators, and $b^i_n$ and $c^i_n$ are ghost oscillators. The matrices $D_{nm}(\gamma )$ and $E_{nm}(\gamma )$ depend on the cyclings and are given by \cite{cg1} \cite{cg2}
\begin{eqnarray}
D_{n0}(\gamma ) & = & \frac{1}{\sqrt{n}}\left[ \gamma (0)\right] ^n\ ,\\
D_{nm}(\gamma ) & = & \left. \sqrt{\frac{m}{n}}\frac{1}{m!}\frac{\partial ^m}
{\partial z^m}\left[ \gamma (z)\right] ^n\right| _{z=0}\ ,\\
D_{00}(\gamma) & = & \left. \frac{1}{2}\ln \left[ \frac{d}{dz}\gamma (z)\right] \right| _{z=0}\ ,\\ 
E_{nm}(\gamma ) & = & \left. \frac{1}{(m+1)!}\frac{\partial ^{m+1}}{\partial z^{m+1}}\left[
\left( \gamma z\right) ^{n+1}\left( \frac{\partial }{\partial z}\gamma z\right) ^{-1}\right]
\right| _{z=0}\ .
\end{eqnarray}
These matrices have the following multiplication properties:
\begin{eqnarray}
\label{eq3}
 & & \sum_{p=1}^\infty D_{np}(\gamma _1)D_{pm}(\gamma _2)+D_{n0}
(\gamma _1)\delta _{0m}+\delta _{0n}D_{0m}(\gamma _2)=D_{nm}(\gamma _1\gamma _2)\ ;\\ 
\label{eq4}
 & & \sum_{t=-1}^1E_{rt}(\gamma _1)E_{ts}(\gamma _2)=E_{rs}(\gamma _1\gamma _2)\ \ ,\ \
r,s,t=-1,0,1\ ;\\
\label{eq5}
 & & E_{rn}(\gamma )=0\ \ ,\ \ r=-1,0,1\ \ ,\ \ n\geq 2\ ,\\
 & & \sum_{p=-1}^\infty E_{np}(\gamma _1)E_{pm}(\gamma _2)=E_{nm}(\gamma _1\gamma _2)\ \ ,\ \ n,m\geq -1\ ;\\
\label{eq6}
 & & \sum_{p=2}^\infty E_{np}(\gamma _1)E_{pm}(\gamma _2)=E_{nm}(\gamma _1\gamma _2)
-\sum_{r,s=-1}^1E_{nr}(\gamma _1)E_{rs}(\gamma _2)\delta _{sm}\ \ ,\ \ n,m\geq 2\ .
\end{eqnarray}
The last term of equation (\ref{eq1}) is the product of $N-3$ ghost oscillators $b^i_n$ ($N$ legs minus arbitrary legs $a$, $b$ and $c$) and is necessary for the vertex to have the correct ghost number. The coefficients $e_n^{ij}$ are given by \cite{ls2}
\begin{equation}
e_n^{ij}=\sum_{m=-1}^\infty k_m^{ij}E_{mn}(\gamma _m^j)\ ,
\end{equation}
where
\begin{eqnarray}
 & & \gamma _{-1}^j=V_j\ ,\\ 
 & & \gamma _0^j=\exp \left( -{\cal L}_0^j\ln a_0^j\right) \exp \left( -\sum_{n=1}^\infty \bar a_n^j{\cal L}_n^j\right) \ ,\\ 
 & & \gamma _p^j=\exp \left( -\sum_{n=p+1}^\infty \bar a_n^j{\cal L}_n^j\right) \ ,\ p\geq 1\ ,
\end{eqnarray}
where ${\cal L}_n^j$ are the generators of the bosonic and ghost parts of the conformal algebra.

\section{Bosonic oscillator part}

Now we shall perform the sewing for the bosonic part of the vertex. The bosonic part of vertex $U_1$ is given by
\begin{equation}
\label{eq2}
V^{\rm osc}_1=\left( \prod_{\scriptstyle i=1\atop \scriptstyle i\neq E}^{N_1}{ }_i\langle 0|\right) \exp \left[ -\frac{1}{2}\sum_{\scriptstyle i,j=1\atop \scriptstyle i\neq j}^{N_1}\sum_{n,m=0}^\infty a^{\mu i}_nD_{nm}\left( \Gamma V^{-1}_iV_j\right) a^j_{m\mu }\right] \ .
\end{equation}
Isolating all terms related to leg $E$, we have the following formula for the modified vertex $V_1^{\rm osc}$:
\begin{eqnarray}
V^{\rm osc}_1 & = & \left( \prod_{\scriptstyle i=1\atop \scriptstyle i\neq E}^{N_1}{ }_i\langle 0|\right) \exp \left[ -\frac{1}{2}\sum_{{\scriptstyle i,j=1\atop \scriptstyle i\neq j}\atop \scriptstyle i,j\neq E}^{N_1}\sum_{n,m=0}^\infty a^{\mu i}_nD_{nm}\left( \Gamma V^{-1}_iV_j\right) a^j_{m\mu }\right] \\
 & & \times { }_E\langle0|\exp \left[ -\sum_{\scriptstyle i=1\atop \scriptstyle i\neq E}^{N_1}\sum_{n,m=0}^\infty a^{\mu i}_nD_{nm}\left( \Gamma V^{-1}_iV_E\right) a^E_{m\mu }\right] \ .
\end{eqnarray}
This may be written in terms of a canonical form (see the Appendix) \cite{cg1}:
\begin{equation}
V^{\rm osc}_1=\left( \prod_{\scriptstyle i=1\atop \scriptstyle i\neq E}^{N_1}{ }_i\langle 0|\right) { }_E\langle 0|\exp \left( -\sum_{n=1}^\infty B^{1\mu }_na^E_{n\mu }\right) \exp (-\phi _1)
\end{equation}
where
\begin{eqnarray}
B^{1\mu }_n & = & \sum_{\scriptstyle i=1\atop \scriptstyle i\neq E}^{N_1}\sum_{n=0}^\infty a^{\mu i}_mD_{mn}\left( \Gamma V^{-1}_iV_E\right) \ ,\\ 
\phi _1 & = & \frac{1}{2}\sum_{{\scriptstyle i,j=1\atop \scriptstyle i\neq j}\atop \scriptstyle i,j\neq E}^{N_1}\sum_{n,m=0}^\infty a^{\mu i}_nD_{nm}\left( \Gamma V^{-1}_iV_j\right) a^j_{m\mu } +\sum_{\scriptstyle i=1\atop \scriptstyle i\neq E}^{N_1}\sum_{n=0}^\infty a^{\mu i}_nD_{n0}\left( \Gamma V^{-1}_iV_E\right) a^E_{0\mu }\ .
\end{eqnarray}

Similarly, considering vertex $V^{{\rm osc}\dagger }_2$, we must change
\begin{equation}
a_m^{\mu F}\rightarrow {a_m^{\mu F}}^\dagger =-a_{-m}^{\mu F}
\end{equation}
so that\footnote{The minus sign in the expression for the Hermitian conjugate is because of the minus sign in the commutation relation of the $a^{\mu i}_n$ oscillators:
\begin{equation}
[a_n^{\mu i},a^{\nu j}_m]=\frac{n}{|n|}\delta _{n,-m}\eta ^{\nu \mu }\delta ^{ij}\ .
\end{equation}}
\begin{eqnarray}
{V^{\rm osc}_2}^\dagger  & = & \left( \prod_{\scriptstyle i=1\atop \scriptstyle i\neq F}^{N_2}{ }_i\langle 0|\right) \exp \left[ -\frac{1}{2}\sum_{{\scriptstyle i,j=1\atop \scriptstyle i\neq j}\atop \scriptstyle i,j\neq F}^{N_2}\sum_{n,m=0}^\infty a^{\mu i}_nD_{nm}\left( \Gamma V^{-1}_iV_j\right) a^j_{m\mu }\right] \nonumber \\
 & & \times \exp \left[ \sum_{\scriptstyle i=1\atop \scriptstyle i\neq F}^{N_2}\sum_{n,m=0}^\infty a^{\mu i}_nD_{nm}\left( \Gamma V^{-1}_iV_E\right) a^F_{-m\mu }\right] |0\rangle _F\ ,
\end{eqnarray}
which can be written
\begin{equation}
V^{{\rm osc}\dagger }_2=\left( \prod_{\scriptstyle i=1\atop \scriptstyle i\neq F}^{N_2}{ }_i\langle 0|\right) \exp \left( -\sum_{n=1}^\infty a^{\mu F}_{-n}A^2_{n\mu }\right) \exp (-\phi _2)|0\rangle _F
\end{equation}
where
\begin{eqnarray}
A^2_{n\mu } & = & -\sum_{\scriptstyle i=1\atop \scriptstyle i\neq F}^{N_2}\sum_{n=0}^\infty D_{nm}\left( \Gamma V^{-1}_FV_i\right) a^i_{m\mu }\ ,\\ 
\phi _2 & = & \frac{1}{2}\sum_{{\scriptstyle i,j=1\atop \scriptstyle i\neq j}\atop \scriptstyle i,j\neq F}^{N_2}\sum_{n,m=0}^\infty a^{\mu i}_nD_{nm}\left( \Gamma V^{-1}_iV_j\right) a^j_{m\mu } +\sum_{\scriptstyle i=1\atop \scriptstyle i\neq F}^{N_2}\sum_{n=0}^\infty a^{\mu i}_nD_{n0}\left( \Gamma V^{-1}_iV_F\right) a^F_{0\mu }\ .
\end{eqnarray}

The oscillator part of the propagator is purely a conformal transformation \cite{pw2}:
\begin{equation}
{\cal P}_{\rm osc}=V^{-1}_EV_F\Gamma \ .
\end{equation}
Using the properties of coherent states (see the Appendix), we may represent this as \cite{cg2}
\begin{equation}
{\cal P}_{\rm osc}=:\exp \left[ \sum_{n,m=0}^\infty a^{\mu E}_{-n}D_{nm}\left( V^{-1}_EV_F\Gamma \right) a^E_{m\mu }-\sum_{n=1}^\infty a^{\mu E}_{-n}a^E_{n\mu }\right] :\ ,
\end{equation}
or, in terms of a canonical form,
\begin{eqnarray}
{\cal P}_{\rm osc} & = & \exp \left(-\sum_{n=1}^\infty a_{-n}^{\mu E}A^3_{n\mu }\right) :\exp \left\{ -\sum_{n,m=1}^\infty a^{\mu E}_{-n}\left[ C^3_{nm}-\delta _{nm}\right] a^E_{m\mu }\right\} :\nonumber \\
 & & \times \exp \left( -\sum_{n=1}^\infty B^{3\mu }_na^E_{n\mu }\right) \exp (-\phi _3)
\end{eqnarray}
with
\begin{eqnarray}
A^3_{n\mu } & = & -D_{n0}\left( V^{-1}_FV_E\Gamma \right) a^E_{0\mu }\ ,\\ 
B^{3\mu }_n & = & a^{\mu E}_0D_{0n}\left( V^{-1}_EV_F\Gamma \right) \ ,\\ 
C^3_{nm} & = & -D_{nm}\left( V^{-1}_EV_F\Gamma \right) \ ,\\ 
\phi _3 & = & a^{\mu E}_0D_{00}\left( V^{-1}_EV_F\Gamma \right) a^E_{0\mu }\ .
\end{eqnarray}

Using now the multiplication rule for two canonical forms (as in the Appendix), we obtain
\begin{eqnarray}
V^{\rm osc}_1{\cal P}_{\rm osc} & = & \left( \prod_{\scriptstyle i=1\atop \scriptstyle i\neq E}^{N_1} { }_i\langle 0|\right) { }_E\langle 0|\exp \left( -\sum_{n=1}^\infty a^{E\mu }_{-n} A^4_{n\mu }\right) \times :\exp \left\{ -\sum_{n,m=1}^\infty a^{\mu E}_{-n} \left[ C^4_{nm}-\delta _{nm}\right] a^E_{m\mu }\right\} : \nonumber \\ 
 & & \times \exp \left( -\sum_{n=1}^\infty B^{4\mu }_na^E_{n\mu }\right) \exp (-\phi _4)
\end{eqnarray}
with
\begin{eqnarray}
A^4_{n\mu } & = & -D_{n0}\left( V^{-1}_FV_E\Gamma \right) a^E_{0\mu }\ ,\\ 
B^{4\mu }_n & = & a^{\mu E}_0D_{0n}\left( V^{-1}_EV_F\Gamma \right) +\sum_{\scriptstyle i=1\atop \scriptstyle i\neq E}^{N_1}\sum_{m=0}^\infty \sum_{p=1}^\infty a^{\mu i}_mD_{mp}\left( \Gamma V^{-1}_iV_E\right) D_{pn}\left( V^{-1}_EV_F\Gamma \right) \ ,\\ 
C^4_{nm} & = & -D_{nm}\left( V^{-1}_EV_F\Gamma \right) \ ,\\ 
\phi _4 & = & \frac{1}{2}\sum_{{\scriptstyle i,j=1\atop \scriptstyle i\neq j}\atop \scriptstyle i,j\neq E}^{N_1}\sum_{n,m=0}^\infty a^{\mu i}_nD_{nm}\left( \Gamma V^{-1}_iV_j\right) a^j_{m\mu }\nonumber \\
 & & +\sum_{\scriptstyle i=1\atop \scriptstyle i\neq E}^{N_1}\sum_{n=0}^\infty a^{\mu i}_nD_{n0}\left( \Gamma V^{-1}_iV_E\right) a^E_{0\mu }+a^{\mu E}_0D_{00}\left( V^{-1}_EV_F\Gamma \right) a^E_{0\mu }\nonumber \\ 
 & & +\sum_{\scriptstyle i=1\atop \scriptstyle i\neq E}^{N_1}\sum_{n=0}^\infty \sum_{n=1}^\infty a^{\mu i}_mD_{nm}\left( \Gamma V^{-1}_iV_E\right) D_{n0}\left( V^{-1}_EV_F\Gamma \right) a^E_{0\mu }\ .
\end{eqnarray}

Because of the vacuum ${}_E\langle 0|$, the first two exponentials do not give any contribution, and so we end up with
\begin{equation}
V^{\rm osc}_1{\cal P}_{\rm osc}=\left( \prod_{\scriptstyle i=1\atop \scriptstyle i\neq E}^{N_1}{ }_i\langle 0|\right) { }_E\langle 0|\exp \left( -\sum_{n=1}^\infty B^{4\mu }_na^E_{n\mu }\right) \exp (-\phi _4)\ .
\end{equation}

Now we make use of property (\ref{eq3}) of the $D_{nm}(\gamma )$ matrices in order to simplify the expressions for the coefficients of $V^{\rm osc}_1{\cal P}_{\rm osc}$. We then have
\begin{eqnarray}
\sum_{n=1}^\infty B^{4\mu }_na^E_{n\mu } & = & \sum_{\scriptstyle i=1\atop \scriptstyle i\neq E}^{N_1}\sum_{m=0}^\infty \sum_{n=1}^\infty a^{\mu i}_mD_{mn}\left( \Gamma V^{-1}_iV_F\Gamma \right) a^E_{n\mu } \nonumber \\ 
 & & +\sum_{n=1}^\infty \left( a^{\mu E}_0+\sum_{\scriptstyle i=1\atop \scriptstyle i\neq E}^{N_1}a^{\mu i}_0\right) D_{0n}\left( V^{-1}_EV_F\Gamma \right) a^E_{n\mu }\ ,\\ 
\phi _4& = & \frac{1}{2}\sum_{{\scriptstyle i=1\atop \scriptstyle i\neq j}\atop \scriptstyle i,j\neq E}^{N_1}\sum_{n,m=0}^\infty a^{\mu i}_nD_{nm}\left( \Gamma V^{-1}_iV_j\right) a^j_{m\mu } +\sum_{\scriptstyle i=1\atop \scriptstyle i\neq E}^{N_1}\sum_{n=0}^\infty a^{\mu i}_nD_{n0}\left( \Gamma V^{-1}_iV_F\Gamma \right) a^E_{0\mu }\nonumber \\ 
 & & +\left( a^{\mu E}_0+\sum_{\scriptstyle i=1\atop \scriptstyle i\neq E}^{N_1}a^{\mu i}_0\right) D_{00}\left( V^{-1}_EV_F\Gamma \right) a^E_{0\mu }\ .
\end{eqnarray}
Since
\begin{equation}
a^{\mu E}_0+\sum_{\scriptstyle i=1\atop \scriptstyle i\neq E}^{N_1}a^{\mu i}_0=\sum_{i=1}^Na^{\mu i}_0=0\ ,
\end{equation}
which is implied by momentum conservation on vertex $V^{\rm osc}_1$, we have the following coefficients for $V^{\rm osc}_1{\cal P}_{\rm osc}$:
\begin{eqnarray}
B^{4\mu }_n & = & \sum_{\scriptstyle i=1\atop \scriptstyle i\neq E}^{N_1}\sum_{m=0}^\infty a^{\mu i}_mD_{mn}\left( \Gamma V^{-1}_iV_F\Gamma \right) \ ,\\ 
\phi _4 & = & \frac{1}{2} \sum_{{\scriptstyle i,j=1\atop \scriptstyle i\neq j}\atop \scriptstyle i,j\neq E}^{N_1}\sum_{n,m=0}^\infty a^{\mu i}_nD_{nm}\left( \Gamma V^{-1}_iV_j\right) a^j_{m\mu } +\sum_{\scriptstyle i=1\atop \scriptstyle i\neq E}^{N_1}\sum_{n=0}^\infty a^{\mu i}_nD_{n0}\left( \Gamma V^{-1}_iV_F\Gamma \right) a^E_{0\mu }\ .
\end{eqnarray}
So, the effect of multiplying $V^{\rm osc}_1$ by ${\cal P}_{\rm osc}$ is just to change the cycling transformations in vertex $V^{\rm osc}_1$ in the following way:
\begin{equation}
V_E\longrightarrow V_F\Gamma \ .
\end{equation}

Now we multiply $V^{\rm osc}_1{\cal P}_{\rm osc}$ by $V^{{\rm osc}\dagger }_2$, obtaining the composite vertex
\begin{equation}
V_c^{\rm gh}=V^{\rm osc}_1{\cal P}_{\rm osc}V^{{\rm osc}\dagger }_2 = \left( \prod_{\scriptstyle i=1\atop \scriptstyle i\neq E,F}^{N_1+N_2}{ }_i\langle 0|\right) { }_E\langle 0|\left( -\sum_{n=1}^\infty a^{\mu E}_{-n}A_{n\mu }\right) \left( -\sum_{n=1}^\infty B^\mu _na^E_{n\mu }\right) \exp (-\phi )|0\rangle _F
\end{equation}
where
\begin{eqnarray}
A_{n\mu } & = & -\sum_{\scriptstyle i=1\atop \scriptstyle i\neq F}^{N_2}\sum_{n=0}^\infty D_{nm}\left( \Gamma V^{-1}_FV_i\right) a^i_{m\mu }\ ,\\ 
B^\mu _n & = & \sum_{\scriptstyle i=1\atop \scriptstyle i\neq E}^{N_1}\sum_{m=0}^\infty a^{\mu i}_mD_{mn}\left( \Gamma V^{-1}_iV_F\Gamma \right) \ ,\\ 
\phi  & = & \frac{1}{2}\sum_{{\scriptstyle i,j=1\atop \scriptstyle i\neq j}\atop \scriptstyle i,j\neq E}^{N_1}\sum_{n,m=0}^\infty a^{\mu i}_nD_{nm}\left( \Gamma V^{-1}_iV_j\right) a^j_{m\mu } +\frac{1}{2}\sum_{{\scriptstyle i,j=1\atop \scriptstyle i\neq j}\atop \scriptstyle i,j\neq F}^{N_2}\sum_{n,m=0}^\infty a^{\mu i}_nD_{nm}\left( \Gamma V^{-1}_iV_j\right) a^j_{m\mu }\nonumber \\
 & & +\sum_{\scriptstyle i=1\atop \scriptstyle i\neq E}^{N_1}\sum_{n=0}^\infty a^{\mu i}_nD_{n0}\left( \Gamma V^{-1}_iV_F\Gamma \right) a^E_{0\mu } +\sum_{\scriptstyle i=1\atop \scriptstyle i\neq F}^{N_2}\sum_{n=0}^\infty a^{\mu i}_nD_{n0}\left( \Gamma V^{-1}_iV_F\right) a^F_{0\mu }\nonumber \\ 
 & & +\sum_{\scriptstyle i=1\atop \scriptstyle i\neq E}^{N_1}\sum_{\scriptstyle i=1\atop \scriptstyle i\neq F}^{N_2}\sum_{n,m=0}^\infty \sum_{p=1}^\infty a^{\mu i}_nD_{np}\left( \Gamma V^{-1}_iV_F\Gamma \right) D_{pm}\left( \Gamma V^{-1}_FV_j\right) a^j_{m\mu }.
\end{eqnarray}

Using the multiplication properties (\ref{eq3}) of the $D_{nm}(\gamma )$ matrices, we get
\begin{eqnarray}
\phi  & = & \frac{1}{2}\sum_{{\scriptstyle i,j=1\atop \scriptstyle i\neq j}\atop \scriptstyle i,j\neq E}^{N_1}\sum_{n,m=0}^\infty a^{\mu i}_nD_{nm}\left( \Gamma V^{-1}_iV_j\right) a^j_{m\mu } +\frac{1}{2}\sum_{{\scriptstyle i,j=1\atop \scriptstyle i\neq j}\atop \scriptstyle i,j\neq F}^{N_2}\sum_{n,m=0}^\infty a^{\mu i}_nD_{nm}\left( \Gamma V^{-1}_iV_j\right) a^j_{m\mu }\nonumber \\
 & & +\sum_{\scriptstyle i=1\atop \scriptstyle i\neq E}^{N_1}\sum_{\scriptstyle i=1\atop \scriptstyle i\neq F}^{N_2}\sum_{n,m=0}^\infty a^{\mu i}_nD_{nm}\left( \Gamma V^{-1}_iV_j\right) a^j_{m\mu } +\sum_{\scriptstyle i=1\atop \scriptstyle i\neq E}^{N_1}\sum_{n=0}^\infty a^{\mu i}_nD_{n0}\left( \Gamma V^{-1}_iV_F\Gamma \right) \left( a^E_{0\mu }+\sum_{\scriptstyle j=1\atop \scriptstyle j\neq F}^{N_2}a^j_{0\mu }\right) \nonumber \\ 
 & & +\sum_{\scriptstyle i=1\atop \scriptstyle i\neq F}^{N_2}\sum_{n=0}^\infty a^{\mu i}_nD_{n0}\left( \Gamma V^{-1}_iV_F\right) \left( a^F_{0\mu }+\sum_{\scriptstyle j=1\atop \scriptstyle j\neq E}^{N_1}a^j_{0\mu }\right) \ .
\end{eqnarray}
Considering now the identification of legs $E$ and $F$ and the momentum conservation for vertices $V^{\rm osc}_1$ and $V^{{\rm osc}\dagger }_2$, this becomes simply
\begin{equation}
\phi =\frac{1}{2}\sum_{{\scriptstyle i,j=1\atop \scriptstyle i\neq j}\atop \scriptstyle i,j\neq E}^{N_1+N_2}\sum_{n,m=0}^\infty a^{\mu i}_nD_{nm}\left( \Gamma V^{-1}_iV_j\right) a^j_{m\mu }\ .
\end{equation}

Because of the two vacua ${ }_E\langle 0|$ and $|0\rangle _F$, the terms $A_{n\mu }$ and $B^\mu _n$ will not give any contribution and we will end up just with the contribution of the phase $\phi $. So we have that
\begin{equation}
V^{\rm osc}_c=\left( \prod_{\scriptstyle i=1\atop \scriptstyle i\neq E,F}^{N_1+N_2}{ }_i\langle 0|\right) \exp \left[ -\frac{1}{2}\sum_{{\scriptstyle i,j=1\atop \scriptstyle i\neq j}\atop \scriptstyle i,j\neq E}^{N_1+N_2}\sum_{n,m=0}^\infty a^{\mu i}_nD_{nm}\left( \Gamma V^{-1}_iV_j\right) a^j_{m\mu }\right] 
\end{equation}
which is the right expression for the composite vertex $V^{\rm osc}_c$.

\subsection{Ghost part}

We will now consider the ghost part of the vertex. The ghost part of a physical vertex $U$ with $N$ legs is given by
\begin{eqnarray}
U^{\rm ghost} & = & \left( \prod_{\scriptstyle i=1\atop \scriptstyle i\neq E}^{N}{ }_i\langle 0|\right) \exp \left[ \sum_{\scriptstyle i,j=1\atop \scriptstyle i\neq j}^N\sum_{n=2}^\infty \sum_{m=-1}^\infty c_n^iE_{nm}\left( \Gamma V^{-1}_iV_j\right) b_m^j\right] \nonumber \\ 
 & & \times \prod_{r=-1}^1\sum_{j=1}^N\sum_{s=-1}^1E_{rs}(V_j)b_s^j\times \prod_{\scriptstyle i=1\atop \scriptstyle i\neq a,b,c}^N\sum_{j=1}^N\sum_{n=-1}^\infty e_n^{ij}b_n^j\ ,
\end{eqnarray}
Making use of the integral of some anticommuting variables $\beta _r$ $(r=-1,0,1)$ and $\eta ^i$ $(i=1,\dots ,N;\ i\neq a,b,c)$, the ghost part of this vertex can be written \cite{cg2}
\begin{eqnarray}
U^{\rm gh}& = & \left( \prod_{i=1}^N{ }_i\langle 0|\right) \int \prod_{r=-1}^1d\beta _r\int \prod_{\scriptstyle i=1\atop \scriptstyle i\neq a,b,c}^Nd\eta ^i \exp \left[ \sum_{\scriptstyle i,j=1\atop \scriptstyle i\neq j}^N\sum_{n=2}^\infty \sum_{m=-1}^\infty c_n^iE_{nm}\left( \Gamma V^{-1}_iV_j\right) b_m^j\right] \nonumber \\
 & & \times \exp \left[ \beta _r\sum_{j=1}^N\sum_{s=-1}^1E_{rs}(V_j)b_s^j\right] \exp \left[ \eta ^i\sum_{j=1}^N\sum_{n=-1}^\infty e_n^{ij}b_n^j\right] \ .
\end{eqnarray}

The ghost part of vertex $U_1$ can be written in a similar way. Isolating all terms related with leg $E$, we obtain
\begin{eqnarray}
U^{\rm gh}_1 & = & \left( \prod_{\scriptstyle i=1\atop \scriptstyle i\neq E}^{N_1}{ }_i\langle 0|\right) { }_E\langle 0|\int \prod_{r=-1}^1d\beta _r\int \prod_{\scriptstyle i=1\atop \scriptstyle i\neq b,c}^{N_1}d\eta ^i \exp \left[ \sum_{{\scriptstyle i,j=1\atop \scriptstyle i\neq j}\atop \scriptstyle i,j\neq E}^{N_1}\sum_{n=2}^\infty \sum_{m=-1}^\infty c_n^iE_{nm}\left( \Gamma V^{-1}_iV_j\right) b_m^j\right] \nonumber \\
 & & \times \exp \left[ \sum_{\scriptstyle i=1\atop \scriptstyle i\neq E}^{N_1}\sum_{n=2}^\infty \sum_{m=-1}^\infty c_n^iE_{nm}\left( \Gamma V^{-1}_iV_E\right) b_m^E\right] \exp \left[ \sum_{\scriptstyle i=1\atop \scriptstyle i\neq E}^{N_1}\sum_{n=2}^\infty \sum_{m=-1}^\infty c_n^EE_{nm}\left( \Gamma V^{-1}_EV_i\right) b_m^i\right] \nonumber \\
 & & \times \exp \left[ \beta _r\sum_{\scriptstyle j=1\atop \scriptstyle j\neq E}^{N_1}\sum_{s=-1}^1E_{rs}(V_j)b_s^j+\beta _r\sum_{s=-1}^1E_{rs}(V_E)b_s^E\right] \nonumber \\
 & & \times \exp \left[ \eta ^i\sum_{\scriptstyle j=1\atop \scriptstyle j\neq E}^{N_1}\sum_{n,m=-1}^\infty e_n^{ij}E_{nm}(V_{0j}^{-1}V_j)b_m^j +\eta ^i\sum_{n,m=-1}^\infty e_n^{iE}E_{nm}(V_{0E}^{-1}V_E)b_m^E\right] \ .
\end{eqnarray}
This may be written in terms of a canonical form\footnote{See the Appendix for the definition of canonical forms for ghosts.} \cite{cg2}:
\begin{equation}
U^{\rm gh}_1 = \left( \prod_{\scriptstyle i=1\atop \scriptstyle i\neq E}^{N_1}{ }_i\langle 0|\right) { }_E\langle 0|\int \prod_{r=-1}^1d\beta _r\int \prod_{\scriptstyle i=1\atop \scriptstyle i\neq b,c}^{N_1}d\eta ^i \exp {\phi _1}\exp \left( \sum_{n=-1}^\infty C_n^1b_n^E\right) \exp \left( \sum_{n=2}^\infty c_n^ED_n^1\right) 
\end{equation}
where
\begin{eqnarray}
\phi _1 & = & \sum_{{\scriptstyle i,j=1\atop \scriptstyle i\neq j}\atop \scriptstyle i,j\neq E}^{N_1}\sum_{n=2}^\infty \sum_{m=-1}^\infty c_n^i E_{nm} \left( \Gamma V^{-1}_iV_j\right) b_m^j +\beta _r \sum_{\scriptstyle j=1\atop \scriptstyle j\neq E}^{N_1} \sum_{s=-1}^1 E_{rs}(V_j)b_s^j\nonumber \\ 
 & & +\eta ^i\sum_{\scriptstyle j=1\atop \scriptstyle j\neq E}^{N_1} \sum_{n,m=-1}^\infty e_n^{ij} E_{nm}(V_{0j}^{-1}V_j)b_m^j\ ,\\ 
C_n^1 & = & \sum_{\scriptstyle i=1\atop \scriptstyle i\neq E}^{N_1} \sum_{m=2}^\infty c_m^i E_{mn}\left( \Gamma V^{-1}_iV_E\right) +\beta _r\sum_{s=-1}^1 E_{rs}(V_E) \delta _{sn} +\sum_{m=-1}^\infty \eta ^i e_m^{iE} E_{mn}(V_{0E}^{-1}V_E)\ ,\\ 
D_n^1 & = & \sum_{\scriptstyle i=1\atop \scriptstyle i\neq E}^{N_1}\sum_{m=-1}^\infty E_{nm}\left( \Gamma V^{-1}_EV_i\right) b_m^i\ .
\end{eqnarray}

Similarly, the physical ghost vertex $U^{{\rm gh}\dagger }_2\equiv C_FU_2^{0{\rm gh}\dagger }$ can be obtained by making the transformations
\begin{equation}
b_n^i\rightarrow b_{-n}^i\ \ ,\ \ c_n^i\rightarrow -c_{-n}^i
\end{equation}
and by using the Hermitian conjugate $|0\rangle _F$ of vacuum ${ }_F\langle 0|$:
\begin{eqnarray}
U^{{\rm gh}\dagger }_2 & = & \left( \prod_{\scriptstyle i=1\atop \scriptstyle i\neq F}^{N_2}{ }_i\langle 0|\right) \int \prod_{r=-1}^1d\beta _r\int \prod_{\scriptstyle i=1\atop \scriptstyle i\neq d,g,h}^{N_2}d{\cal M}^i \exp \left[ \sum_{{\scriptstyle i,j=1\atop \scriptstyle i\neq j}\atop \scriptstyle i,j\neq F}^{N_2}\sum_{n=2}^\infty \sum_{m=-1}^\infty c_n^iE_{nm}\left( \Gamma V^{-1}_iV_j\right) b_m^j\right] \nonumber \\
 & & \times \exp \left[ \sum_{\scriptstyle i=1\atop \scriptstyle i\neq F}^{N_2}\sum_{n=2}^\infty \sum_{m=2}^\infty b_{-m}^Fc_n^iE_{nm}\left( \Gamma V^{-1}_iV_F\right) \right] \nonumber \\ 
 & & \times \exp \left[ -\sum_{\scriptstyle i=1\atop \scriptstyle i\neq F}^{N_2}\sum_{n=2}^\infty \sum_{m=-1}^\infty E_{nm}\left( \Gamma V^{-1}_FV_i\right) b_m^ic_{-n}^F\right] \exp \left[ \beta _r\sum_{\scriptstyle j=1\atop \scriptstyle j\neq F}^{N_2}\sum_{s=-1}^1E_{rs}(V_j)b_s^j\right] \nonumber \\ 
 & & \times \exp \left[ {\cal M}^i\sum_{\scriptstyle j=1\atop \scriptstyle j\neq E}^{N_2}\sum_{n,m=-1}^\infty e_n^{ij}E_{nm}(V_{0j}^{-1}V_j)b_m^j +\sum_{n=2}^\infty \sum_{m=-1}^\infty b_{-m}^F{\cal M}_ie_n^{iF}E_{nm}(V_{0F}^{-1}V_F)\right] |0\rangle _F
\end{eqnarray}
where some terms obtained by following the procedure described above have been annihilated by the Hermitian conjugate vacuum $|0\rangle _F$.

This vertex can be written as
\begin{equation}
U^{{\rm gh}\dagger }_2 = \left( \prod_{\scriptstyle i=1\atop \scriptstyle i\neq F}^{N_2}{ }_i\langle 0|\right) \int \prod_{r=-1}^1d\beta _r\int \prod_{\scriptstyle i=1\atop \scriptstyle i\neq d,g,h}^{N_2}d{\cal M}^i \exp {\phi _2}\exp \left( \sum_{n=2}^\infty A_n^2b_{-n}^F\right) \exp \left( \sum_{n=-1}^\infty c_{-n}^FB_n^2\right) |0\rangle _F
\end{equation}
where
\begin{eqnarray}
\phi _2 & = & \sum_{{\scriptstyle i,j=1\atop \scriptstyle i\neq j}\atop \scriptstyle i,j\neq F}^{N_2}\sum_{n=2}^\infty \sum_{m=-1}^\infty c_n^iE_{nm}\left( \Gamma V^{-1}_iV_j\right) b_m^j +\beta _r\sum_{\scriptstyle j=1\atop \scriptstyle j\neq F}^{N_2}\sum_{s=-1}^1E_{rs}(V_j)b_s^j\nonumber \\
 & & +{\cal M}^i\sum_{\scriptstyle j=1\atop \scriptstyle j\neq F}^{N_2}\sum_{n,m=-1}^\infty e_n^{ij}E_{nm}(V_{0j}^{-1}V_j)b_m^j\ ,\\
A_n^2 & = & -\sum_{\scriptstyle i=1\atop \scriptstyle i\neq F}^{N_2}\sum_{m=2}^\infty c_m^iE_{mn}\left( \Gamma V^{-1}_iV_F\right) -\sum_{m=-1}^\infty {\cal M}^ie_m^{iF}E_{mn}(V_{0F}^{-1}V_F)\ ,\\ 
B_n^2 & = & \sum_{\scriptstyle i=1\atop \scriptstyle i\neq F}^{N_2}\sum_{p=2}^\infty \sum_{m=-1}^\infty \delta_{np}E_{pm}\left( \Gamma V^{-1}_FV_i\right) b_m^i\ .
\end{eqnarray}

The ghost part of the propagator is given by
\begin{equation}
{\cal P}=V_E^{-1}V_F\Gamma \ .
\end{equation}
Being a conformal transformation, can be written in terms of the following canonical form \cite{cg2}:
\begin{eqnarray}
{\cal P}_{\rm gh} & = & :\exp \left[ \sum_{n,m=-1}^\infty c_{-n}^E\left( E_{nm}^3-\delta _{nm}\right) b_m^E\right] \exp \left( \sum_{n=2}^\infty \sum_{r=-1}^1c_n^EG_{nr}^3b_r^E\right) \nonumber \\ 
 & & \times \exp \left[ \sum_{n,m=2}^\infty b_{-n}^E\left( F_{nm}^3-\delta _{nm}\right) c_m^E\right] :
\end{eqnarray}
with
\begin{eqnarray}
 & & E_{nm}^3 = E_{nm}\left( V_E^{-1} V_F\Gamma \right) \ ,\\ 
 & & G_{nr}^3 = -\sum_{s=-1}^1 E_{ns} \left( V_F^{-1} V_E\Gamma \right) E_{sr}\left( \Gamma V_E^{-1} V_F\Gamma \right) \ ,\\ 
 & & F_{nm}^3 = E_{nm}\left( V_E^{-1} V_F\Gamma \right) \ .
\end{eqnarray}

Multiplying now $U^{\rm gh}_1$ by ${\cal P}_{\rm gh}$, we obtain
\begin{eqnarray}
U^{\rm gh}_1{\cal P}_{\rm gh} & = & \left( \prod_{\scriptstyle i=1\atop \scriptstyle i\neq E}^{N_1}{ }_i\langle 0|\right) { }_E\langle 0|\int \prod_{r=-1}^1d\beta _r\int \prod_{\scriptstyle i=1\atop \scriptstyle i\neq b,c}^{N_1}d\eta ^i \exp {\phi _4}:\exp \left( \sum_{n=2}^\infty \sum_{r=-1}^1c_n^EG_{nr}^4b_r^E\right) :\nonumber \\
 & & \times \exp \left( \sum_{n=-1}^\infty C_n^4b_n^E\right) \exp \left( \sum_{n=2}^\infty c_n^ED_n^4\right) 
\end{eqnarray}
where
\begin{eqnarray}
\phi _4 & = & \phi _1\ ,\\ 
G_{nr}^4 & = & G_{nr}^3\ ,\\ 
C_n^4 & = & \sum_{\scriptstyle i=1\atop \scriptstyle i \neq E}^{N_1} \sum_{m=2}^\infty c_m^i E_{mn} \left( \Gamma V^{-1}_i V_F \Gamma \right) +\beta _r \sum_{s=-1}^1 E_{rs} \left( V_F\Gamma \right) \delta {sn} \eta _i \sum_{m=-1}^\infty e_m^{iE} E_{mn} \left( V_{0E}^{-1} V_F\Gamma \right) \ ,\\ 
D_n^4 & = & \sum_{\scriptstyle i=1\atop \scriptstyle i \neq E}^{N_1} \sum_{m=-1}^\infty E_{nm} \left( V^{-1}_F V_i\right) b_m^i -\sum_{\scriptstyle i=1\atop \scriptstyle i\neq E}^{N_1} \sum_{r,s=-1}^1 E_{nr} \left( V^{-1}_FV_E \Gamma \right) E_{rs} \left( \Gamma V^{-1}_E V_i\right) b_s^i
\end{eqnarray}
and some of the terms have been annihilated by the vacuum ${ }_E\langle 0|$. In the expressions above, we have used multiplication properties (\ref{eq4}-\ref{eq6}) of matrices $E_{nm}(\gamma )$.

The combination of the second and fourth terms in the expression above yields
\begin{eqnarray}
 & & :\exp \left( \sum_{n=2}^\infty \sum_{r=-1}^1c_n^EG_{nr}^4b_r^E\right) :\exp \left( \sum_{n=2}^\infty c_n^ED_n^4\right) = \nonumber \\ 
 & & \ \ \ \ \ \ \ \ \ \ -\sum_{n=2}^\infty \sum_{r,s=-1}^1c_n^EE_{ns}\left( V_F^{-1}V_E\Gamma \right) \left[ E_{sr}\left( \Gamma V_E^{-1}V_F\Gamma \right) b_r^E+\sum_{\scriptstyle i=1\atop \scriptstyle i\neq E}^{N_1}E_{sr}\left( \Gamma V_E^{-1}V_i\right) b_r^i\right] \ .
\end{eqnarray}
Using multiplication property (\ref{eq4}) we can write this as
\begin{equation}
-\sum_{n=2}^\infty \sum_{r,s,t=-1}^1c_n^EE_{ns}\left( V_F^{-1}V_E\Gamma \right) E_{st}\left( \Gamma V_E^{-1}\right) \left[ E_{tr}\left( V_F\Gamma \right) b_r^E+\sum_{\scriptstyle i=1\atop \scriptstyle i\neq E}^{N_1}E_{tr}\left( V_i\right) b_r^i\right] 
\end{equation}
what is zero due to the overlap identities of $U_1^{\rm gh}{\cal P}_{\rm gh}$.

We can now rewrite the combination $U_1^{\rm gh}{\cal P}_{\rm gh}$ as \cite{cg2}
\begin{equation}
U^{\rm gh}_1{\cal P}_{\rm gh} = \left( \prod_{\scriptstyle i=1\atop \scriptstyle i\neq E}^{N_1}{ }_i\langle 0|\right) { }_E\langle 0|\int \prod_{r=-1}^1d\beta _r\int \prod_{\scriptstyle i=1\atop \scriptstyle i\neq b,c}^{N_1}d\eta ^i \exp {\phi _5}\exp \left( \sum_{n=-1}^\infty C_n^5b_n^E\right) \exp \left( \sum_{n=2}^\infty c_n^ED_n^5\right) \end{equation}
where now
\begin{eqnarray}
\phi _5 & = & \sum_{{\scriptstyle i,j=1\atop \scriptstyle i \neq j}\atop \scriptstyle i,j \neq E}^{N_1} \sum_{n=2}^\infty \sum_{m=-1}^\infty c_n^i E_{nm} \left( \Gamma V^{-1}_i V_j\right) b_m^j +\beta _r \sum_{\scriptstyle j=1\atop \scriptstyle j\neq E}^{N_1} \sum_{s=-1}^1 E_{rs}(V_j) b_s^j\nonumber \\ 
 & & +\eta ^i \sum_{\scriptstyle j=1\atop \scriptstyle j\neq E}^{N_1} \sum_{n,m=-1}^\infty e_n^{ij} E_{nm} (V_{0j}^{-1} V_j) b_m^j\ ,\\ 
C_n^5 & = & \sum_{\scriptstyle i=1\atop \scriptstyle i \neq E}^{N_1} \sum_{m=2}^\infty c_m^i E_{mn} \left( \Gamma V^{-1}_i V_F \Gamma \right) +\beta _r \sum_{s=-1}^1 E_{rs} \left( V_F \Gamma \right) \delta _{sn} +\eta ^i \sum_{m=-1}^\infty e_m^{iE} E_{mn} \left( V_{0E}^{-1} V_F\Gamma \right) \ ,\\ 
D_n^5 & = & \sum_{\scriptstyle i=1\atop \scriptstyle i \neq E}^{N_1} \sum_{m=-1}^\infty E_{nm} \left( V^{-1}_F V_i\right) b_m^i\ .
\end{eqnarray}
So the effect of inserting the propagator into $U_1G$ is to make the following change:
\begin{equation}
V_E\rightarrow V_E{\cal P}=V_F\Gamma 
\end{equation}
what was expected considering the bosonic oscillator case.

Now we multiply $U^{\rm gh}_1{\cal P}_{\rm gh}$ by $U^{{\rm gh}\dagger }_2$ in order to obtain the composite vertex $U_c^{\rm gh}$. The only remaining term which is not annihilated by the two vacua associated to legs $E$ and $F$ is the resulting phase $\phi $, so that the result obtained is\footnote{We write $U^{\rm gh}_c\equiv U^{\rm gh}_1{\cal P}_{\rm gh}U_2^{{\rm gh}\dagger }$.}
\begin{equation}
U_c^{\rm gh}=\left( \prod_{\scriptstyle i=1\atop \scriptstyle i\neq E,F}^{N_1+N_2}{ }_i\langle 0|\right) \int \prod_{r=-1}^1d\beta _r\int \prod_{\scriptstyle i=1\atop \scriptstyle i\neq b,c}^{N_1}d\eta ^i\int \prod_{\scriptstyle i=1\atop \scriptstyle i\neq d,g,h}^{N_2}d{\cal M}^i\exp \phi 
\end{equation}
where
\begin{eqnarray}
\phi  & = & \sum_{{\scriptstyle i,j=1\atop \scriptstyle i \neq j}\atop \scriptstyle i,j\neq E}^{N_1} \sum_{n=2}^\infty \sum_{m=-1}^\infty c_n^i E_{nm}\left( \Gamma V^{-1}_iV_j \right) b_m^j +\sum_{{\scriptstyle i,j=1\atop \scriptstyle i\neq j}\atop \scriptstyle i,j\neq F}^{N_2} \sum_{n=2}^\infty \sum_{m=-1}^\infty c_n^i E_{nm}\left( \Gamma V^{-1}_i V_j \right) b_m^j\nonumber \\
 & & +\sum_{\scriptstyle i=1\atop \scriptstyle i\neq E}^{N_1} \sum_{\scriptstyle j=1\atop \scriptstyle j \neq F}^{N_2} \sum_{n=2}^\infty \sum_{m=-1}^\infty c_n^i E_{nm}\left( \Gamma V^{-1}_i V_j\right) b_m^j +\sum_{\scriptstyle j=1\atop \scriptstyle j \neq E}^{N_1} \sum_{\scriptstyle i=1\atop \scriptstyle i\neq F}^{N_2} \sum_{n=2}^\infty \sum_{m=-1}^\infty c_n^i E_{nm}\left( \Gamma V^{-1}_i V_j\right) b_m^j\nonumber \\ 
 & & +\beta _r \sum_{\scriptstyle j=1\atop \scriptstyle j\neq E}^{N_1} \sum_{s=-1}^1 E_{rs}(V_j) b_s^j +\beta _r \sum_{\scriptstyle i=1\atop \scriptstyle i \neq E}^{N_2} \sum_{s=-1}^1 E_{rs}(V_i) b_s^i\nonumber \\ 
 & & +\eta ^i \sum_{\scriptstyle j=1\atop \scriptstyle j\neq E}^{N_1} \sum_{n=-1}^\infty e_n^{ij} E_{nm} (V_{0j}^{-1} V_j) b_m^j +\eta ^i \sum_{\scriptstyle j=1\atop \scriptstyle j \neq F}^{N_2} \sum_{n,m=-1}^\infty e_m^{iE} E_{mn} \left( V_{0E}^{-1} V_j\right) b_n^j\nonumber \\ 
 & & +{\cal M}^i \sum_{\scriptstyle j=1 \atop \scriptstyle j \neq F}^{N_2} \sum_{n=-1}^\infty e_n^{ij} E_{nm}(V_{0j}^{-1} V_j) b_m^j +{\cal M}^i \sum_{\scriptstyle j=1\atop \scriptstyle j \neq E}^{N_1} \sum_{n,m=-1}^\infty e_m^{iF} E_{mn} \left( V_{0F}^{-1} V_j\right) b_n^j
\end{eqnarray}
where we have set some terms to zero using property (\ref{eq5}) and the overlap identities for vertices $U_1^{\rm gh}$ and $U_2^{{\rm gh}\dagger }$.

Performing the integrations of the anticommuting variables introduced before, we then obtain
\begin{eqnarray}
\lefteqn{U^{\rm gh}_c=\left( \prod_{\scriptstyle i=1\atop \scriptstyle i\neq E,F}^{N_1+N_2}{ }_i\langle 0|\right) \exp \left[ \sum_{{\scriptstyle i,j=1\atop \scriptstyle i\neq j}\atop \scriptstyle i,j\neq E,F}^{N_1+N_2}\sum_{n=2}^\infty \sum_{m=-1}^\infty c_n^iE_{nm}\left( \Gamma V^{-1}_iV_j\right) b_m^j\right] \times \prod_{r=-1}^1\sum_{\scriptstyle i=1\atop \scriptstyle i\neq E,F}^{N_1+N_2}\sum_{s=-1}^1E_{rs}(V_i)b_s^i}\nonumber \\ 
 & & \times \prod_{\scriptstyle i=1\atop \scriptstyle i\neq b,c}^{N_1}\left[ \sum_{\scriptstyle j=1\atop \scriptstyle j\neq E}^{N_1}\sum_{n,m=-1}^\infty e_n^{ij}E_{nm}(V_{0j}^{-1}V_j)b_m^j+\sum_{\scriptstyle j=1\atop \scriptstyle j\neq F}^{N_2}\sum_{n,m=-1}^\infty e_m^{iE}E_{mn}\left( V_{0E}^{-1}V_j\right) b_n^j\right] \nonumber \\ 
 & & \times \! \! \prod_{\scriptstyle i=1\atop \scriptstyle i\neq d,g,h}^{N_2}\left[ \sum_{\scriptstyle j=1\atop \scriptstyle j\neq E}^{N_1}\sum_{n,m=-1}^\infty e_m^{iF}E_{mn}\left( V_{0F}^{-1}V_j\right) b_n^j+\sum_{\scriptstyle j=1\atop \scriptstyle j\neq F}^{N_2}\sum_{n,m=-1}^\infty e_n^{ij}E_{nm}(V_{0j}^{-1}V_j)b_m^j\right] .
\end{eqnarray}
This expression for the composite vertex matches exactly the one obtained using the overlap identities \cite{ls1}.

\section{Conclusions}

Using explicit sewing techniques, we have sewn two vertices together in order to become a composite vertex. The calculations have been done with the correct ghost numbers for each vertex and the result has both BRST invariance and the correct ghost counting. It verifies the results obtained using overlap identities and the Group Theoretic method for String Theory.

\vskip 0.5 cm

\noindent {\large \bf Acknowledgements}

\vskip 0.3 cm

This work has been done in the Department of Mathematics, King's College London, University of London, under the supervision of Professor P. C. West, and finished at the State University of Santa Catarina (UDESC). The author wishes to thank Professor West for suggesting, guiding and correcting this work. My gratitude to CAPES, Brazil, for financial support. G.E.D.!

\appendix

\section{Coherent states and canonical forms}

In this appendix we present the concepts of coherent states and canonical forms for bosonic and ghost oscillators \cite{cg1} \cite{cg2}.

\subsection{Coherent states}

Given operators $a_n$ with commutation relations
\begin{equation}
[a_n,a_m^\dagger ]=n\delta_{nm}\ ,
\end{equation}
a {\sl coherent state} $|f\rangle $ is defined by\footnote{We will not be writing the index $n$ of $a_n$ until it becomes necessary.}
\begin{equation}
|f\rangle ={\rm e}^{fa^\dagger }|0\rangle \ .
\end{equation}
Some of the properties of coherent states are:
\vskip 0.5 cm

\noindent{\bf 1)} $a|f\rangle =f|f\rangle $ so that ${\rm e}^{ga}|f\rangle ={\rm e}^{gf}|f\rangle $ ;
\vskip 0.3 cm

\noindent{\bf 2)} ${\rm e}^{ga^\dagger }=|f+g\rangle $ ;
\vskip 0.3 cm
\noindent{\bf 3)} $\langle f|g\rangle ={\rm e}^{f^\dagger g}$ ;
\vskip 0.3 cm
\noindent{\bf 4)} $x^{na^\dagger a}|f\rangle =|x^nf\rangle $ ;
\vskip 0.3 cm
\noindent{\bf 5)} $1=\frac{1}{\pi }\int d({\rm Re }f)d({\rm Im}f){\rm e}^{-|f|^2}|f\rangle \langle f|$ ;
\vskip 0.3 cm
\noindent{\bf 6)} Defining the tensor product of coherent states as
\begin{equation}
|f_1,f_2,\dots ,f_n,\dots \rangle =\prod_{n=1}^\infty {\rm e}^{f_na_n^\dagger }|0\rangle \ ,
\end{equation}
we have
\begin{equation}
:\exp \left[ \sum_{n,m=1}^\infty a_n^\dagger (C_{nm}-\delta _{nm})a_m\right] :|f_1,f_2,\dots ,f_n,\dots \rangle = |\sum_{n=1}^\infty C_{1m}f_m,C_{2m}f_m,\dots ,C_{nm}f_m,\dots \rangle \ .
\end{equation}
Symbolically, we can write this property as
\begin{equation}
:{\rm e}^{a^\dagger (C-1)a}:|f\rangle =|Cf\rangle \ .
\end{equation}
Application: since $x^{a^\dagger a}|f\rangle =|xf\rangle $ and $:{\rm e}^{a^\dagger (x-1)a}:|f\rangle =|xf\rangle $, we have
\begin{equation}
x^{a^\dagger a}=:{\rm e}^{a^\dagger (x-1)a}:
\end{equation}
when applied to coherent states.

\subsection{Canonical forms}

There will be two definitions here for canonical forms. In terms of bosonic oscillators \cite{cg1} $\alpha _n$, $[\alpha _n,\alpha _m]=\delta _{n,-m}$, a canonical form ${\cal O}$ is any operator which can be written as
\begin{equation}
{\cal O} = \exp \left( -\sum_{n=1}^\infty \alpha _{-n}A_n\right) :\exp \left[ \sum_{n,m=1}^\infty \alpha _{-n}\left( C_{nm}-\delta _{nm}\right) \alpha _m\right] : \exp \left( -\sum_{n=1}^\infty B_n\alpha _n\right) {\rm e}^{-\phi }
\end{equation}
where $\phi $, $A_n$ and $B_n$ do not depend on $\alpha _n$ or $\alpha _{-n}$.
The product of two canonical forms ${\cal O}_1$ and ${\cal O}_2$ is again a canonical form ${\cal O}$ with
\begin{eqnarray}
 & & A_n=A_n^1+\sum_{m=1}^\infty C_{nm}^1A_m^2\ ,\\
 & & B_n=B_n^2+\sum_{m=1}^\infty B_m^1C_{mn}^2\ ,\\
 & & C_{nm}=\sum_{p=1}^\infty C_{np}^1C_{pm}^2\ ,\\
 & & \phi =\phi _1+\phi _2+\sum_{n=1}^\infty B_n^1A_n^2\ .
\end{eqnarray}
In terms of the ghost oscillators \cite{cg2}, a canonical form ${\cal C}$ can be defined as any operator that can be written as
\begin{eqnarray}
{\cal C} & = & {\rm e}^\phi \exp \left( \sum_{n=2}^\infty A_nb_{-n}\right) \exp \left( \sum_{n=-1}^\infty B_nc_{-n}\right) \nonumber \\ 
 & & \times :\exp \left[ \sum_{n,m=-1}^\infty c_{-n}\left( E_{nm}-\delta _{nm}\right) b_m\right] \exp \left( \sum_{n=2}^\infty \sum_{r=-1}^1c_nG_{nr}b_r\right) \nonumber \\
 & & \times \exp \left[ \sum_{n,m=2}^\infty b_{-n}\left( F_{nm}-\delta _{nm}\right) c_m\right] :\exp \left( \sum_{n=-1}^\infty C_nb_n\right) \exp \left( \sum_{n=2}^\infty D_nc_n\right) \ .
\end{eqnarray}
Again, the product of two canonical forms ${\cal C}_1$ and ${\cal C}_2$ will be a canonical form ${\cal C}$ with coefficients
\begin{eqnarray}
 & & \phi =\phi _1+\phi _2-\sum_{n=-1}^\infty C_n^1B_n^2-\sum_{n=2}^\infty D_n^1A_n^2+\sum_{n=2}^\infty \sum_{r=-1}^1A_n^2G_{nr}^1B^2_r\ ,\\ 
 & & A_n=A^1_n+\sum_{n=2}^\infty A_m^2{F_{mn}^1}^T\ ,\\ 
 & & B_n=B_n^1+\sum_{n=-1}^\infty B_m^2{E_{mn}^1}^T\ ,\\
 & & E_{nm}=\sum_{p=-1}^\infty E_{np}^1E_{pm}^2\ ,\\ 
 & & F_{nm}=\sum_{p=2}^\infty F_{np}^1F_{pm}^2\ ,\\ 
 & & G_{nr}=G_{nr}^2+\sum_{m=2}^\infty \sum_{s=-1}^1F_{nm}^2G_{ms}^1E_{sr}^2\ ,\\
 & & C_n=C_n^2-\sum_{m=2}^\infty \sum_{r=-1}^1A_m^2G_{mr}^1E_{rn}^2+\sum_{m=-1}^\infty C_m^1E_{mn}^2\ ,\\
 & & D_n=D_n^2+\sum_{m=2}^\infty \sum_{r=-1}^1F_{nm}^2G_{mr}^1B_r^2+\sum_{m=2}^\infty D_m^1F_{mn}^2\ .
\end{eqnarray}
where $A^T$ means the transpose matrix of $A$.

\end{document}